\documentclass[prc,twocolumn,showpacs,showkeys,nofootinbib,superscriptaddress]{revtex4}
\usepackage{graphicx}
\usepackage{dcolumn}
\usepackage{epsfig}
\usepackage{bm}

\newcommand{\eq}{\begin{eqnarray}}
\newcommand{\en}{\end{eqnarray}}

\def\rpm{R_p \hspace{-0.8em}/\;\:}
\def\rp{$R_p\hspace{-1em}/\ \ $}
\def\rpp{$R_p\hspace{-0.8em}/\ \ $}
\def\dbd{$0\nu\beta\beta$}

\newcommand{\ba}[1]{\begin{eqnarray} \label{(#1)}}
\newcommand{\ea}{\end{eqnarray}}
\newcommand{\nn}{\nonumber}

\begin{document}

\title{Pion dominance in \rpp SUSY induced neutrinoless double beta decay}

\author{Amand Faessler}
\affiliation{Institute f\"{u}r Theoretische Physik der Universit\"{a}t
T\"{u}bingen, D-72076 T\"{u}bingen, Germany}
\author{Thomas Gutsche}
\affiliation{Institute f\"{u}r Theoretische Physik der Universit\"{a}t
T\"{u}bingen, D-72076 T\"{u}bingen, Germany}
\author{Sergey Kovalenko}
\affiliation{Centro de Estudios Subat\'omicos(CES),
Universidad T\'ecnica Federico Santa Mar\'\i a, \\
Casilla 110-V, Valpara\'\i so, Chile}
\author{Fedor \v Simkovic}
\altaffiliation{On  leave of absence from Department of Nuclear
Physics, Comenius University, Mlynsk\'a dolina F1, SK--842 15
Bratislava, Slovakia}
\affiliation{Institute f\"{u}r Theoretische Physik der Universit\"{a}t
T\"{u}bingen, D-72076 T\"{u}bingen, Germany}

\date{\today}

\begin{abstract}
At the quark level there are basically two types of
contributions of R-parity violating SUSY (\rp SUSY)
to neutrinoless double beta decay: the
short-range contribution involving only heavy virtual superpartners
and the long-range one with the virtual squark and neutrino.
Hadronization of the effective operators, corresponding to these two
types of contributions, may in general involve virtual
pions in addition to close on-mass-shell nucleons. From the previous studies it is known that
the short-range contribution is dominated by the pion exchange.
In the present paper we show that this is also true for
the long-range \rp SUSY contribution. Therefore, we conclude that
the \rp SUSY contributes to the neutrinoless double beta decay
dominantly via charged pion exchange between the decaying nucleons.
\end{abstract}



\maketitle

\newpage

\section{Introduction}

The nuclear neutrinoless double beta decay (\dbd)
is a process known for 70 years, which has been searched for,
but not yet seen. The $0\nu\beta\beta$-decay plays a special role
among exotic processes for the following two reasons. First, it is
intimately related to the nature of the neutrino\cite{relation}, since
it is able to probe whether the neutrino is a  Dirac or a Majorana
particle. Second, the modern \dbd-experiments have reached
unprecedented sensitivity. The most stringent lower bound
on the half-life of the $0\nu\beta\beta$-decay  was measured for $^{76}{Ge}$
in the Heidelberg-Moscow experiment \cite{ge76}:
\begin{equation}\label{H-M}
T^{0\nu}_{1/2}(^{76}Ge) \geq 1.9\times 10^{25}\mbox{yrs}.
\end{equation}
In the near future this limit is expected to be improved
in the GERDA experiment \cite{gerda} by 1-2 orders of magnitude.
For the $0\nu\beta\beta$-decays of $^{100}Mo$
and  $^{130}Te$ in the two running experiments,
NEMO3 \cite{mo100} and CUORICINO \cite{te130},
the following sensitivities have been achieved:
\begin{eqnarray}\label{nemo-cuore}
T^{0\nu}_{1/2}(^{100} \rm{Mo})  \geq 5.8 ~ 10^{23}\, years\nonumber\\
T^{0\nu}_{1/2}(^{130} \rm{Te})  \geq 3.0 ~ 10^{24}\, years.
\label{eq:4}
\end{eqnarray}

The extraordinary sensitivity of \dbd-experiments makes them also a unique
laboratory tool to probe physics beyond the standard model (SM)
with possible lepton number violation (LNV) underlying. Of particular interest in this context are supersymmetric models with R-parity violation (\rp SUSY)
containing the LNV interactions necessary to generate the Majorana mass for neutrinos and
to induce the \dbd-decay. There is a wealth of literature on the \rp SUSY mechanisms of the
\dbd-decay \cite{Moh86,Ver87,dbd-gluino-neutralino,dbd-gluino-neutralino1,Fedor-Wodecki,ramsey,FKS98b,bivalle,HKK:96,Pes}
where  the corresponding quark-level LNV interactions as well as hadronic and nuclear structure aspects relevant for these mechanisms are studied.
At the quark-level there are basically two types of \rp SUSY mechanisms:  the short-range mechanism with the exchange of heavy superpartners
\cite{Moh86,Ver87,dbd-gluino-neutralino,dbd-gluino-neutralino1,Fedor-Wodecki,ramsey}
and the long-range
mechanism involving both the exchange of heavy squarks and the light neutrino
\cite{FKS98b,bivalle,HKK:96,Pes}, which we call squark-neutrino mechanism. For the latter case, due to the chiral structure
of the \rp SUSY interactions, the amplitude of \dbd-decay does not vanish in the limit of zero neutrino mass in contrast to from the ordinary Majorana neutrino
exchange mechanism proportional to the light neutrino mass.
Instead, the squark-neutrino mechanism is roughly proportional to the momentum of the virtual neutrino, which is of the order of the Fermi momentum of the nucleons inside the nucleus
with $p_F\approx 100$MeV. This is a manifestation of the fact that the LNV necessary for \dbd-decay is supplied by the \rp SUSY interactions instead of the Majorana neutrino mass term
and, therefore, this mechanism is not suppressed by the small neutrino mass.

In the calculation of the \dbd-decay amplitude one has to hadronize the quark-level LNV operators representing them in terms of the interpolating nucleon fields.
As is known \cite{dbd-gluino-neutralino1}, this  can be done in two ways: the quark fields can be directly imbedded into the interpolating two nucleon fields (the two nucleon mode)
or via an intermediate step when one quark-antiquark pair is associated with the charged pion field coupled to the nucleons (pion mode).
In Refs. \cite{dbd-gluino-neutralino1,ramsey}
it was shown that the pion mode of hadronization dominates over the two nucleon one for the case of the short-range mechanism.
In the present paper we demonstrate that the same conclusion is valid for the long-range squark-neutrino mechanism.  Therefore, in all cases the dominant contribution of the \rp SUSY
to the \dbd-decay is realized via the pion mode of hadronization. For the calculation of the nuclear
matrix elements (NMEs) of the corresponding transition operators we apply
the proton-neutron Quasiparticle Random Phase Approximation (pn-QRPA) \cite{schwieger}.

The paper is organized as follows. In Sect. II we introduce the quark-level process
related to the squark-neutrino mechanism of the \dbd. In Sect. III we discuss the two nucleon
and the pion  modes of hadronization of the effective quark-level Lagrangian. The
\dbd   \mbox{ NMEs} of the corresponding transition operators are presented in Sect. IV.
For the three nuclei of experimental interest the values of these NMEs are calculated
in the QRPA. They are discussed in Sect. V.  We summarize and discuss our results
in Section VI.

\begin{figure*}[tb]
    \epsfxsize=0.90\textwidth
    \epsffile{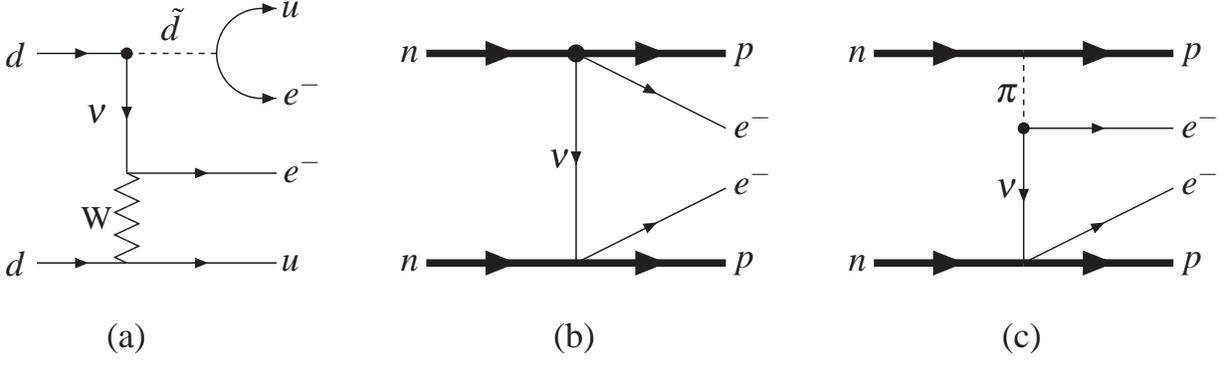}
\caption{
Diagrams describing the squark-neutrino mechanism of
\dbd-decay (a) at the quark-level;  at the nucleon level
with hadronization in the (b) 2N-mode and (c) the pion-mode.
The dark blobs denote the LNV vertices generated by \rp SUSY trilinear interactions.
}
\label{fig:setesmJ}
\end{figure*}

\section{Squark-neutrino mechanism of \dbd-decay}
\label{s2}
Lepton number violation  in the fermions sector is generated within \rp SUSY models
by the following terms of the superpotential:

%
\begin{eqnarray}\label{superpotential}
  W_{\rpm} =     \frac{1}{2} \lambda_{ijk} L_i L_j \bar E_k
    + \lambda'_{ijk} L_i Q_j \bar D_{k} + \kappa^i L_i H_2,
\end{eqnarray}
where  $L$ and $Q$ are the left-handed lepton and quark superfield
$SU(2)$  doublets, while $\bar E$, $\bar U$ and $\bar D$ denote the
right-handed lepton, up-quark and down-quark $SU(2)$ singlets,
respectively.

The mixing between the scalar superpartners $\tilde q_{L,R}$ of the left and
right-handed quarks $q_{L,R}$ plays  the crucial role for the case of
the long-range \rp SUSY squark-neutrino (SQN) mechanism.
This effect occurs due to the non-diagonality of the squark mass matrix. For the down
squarks of each generation it takes the form (e.g. \cite{RPV-reviev})
\begin{eqnarray}\label{e4}
{\cal M}^2_{\tilde d} &= \mbox{$ \left( \begin{array}{cc}
m^2_{\tilde{d}_L} + m_d^2 - 0.42 D_Z & - m_d (A_d + \mu \tan\! \beta) \\
- m_d (A_d + \mu \tan\! \beta) & m^2_{\tilde{d}_R} + m_d^2 - 0.08 D_Z
\end{array} \right) $}.\nonumber \\
\end{eqnarray}
Here,  $ d = d,  s,  b$  and $\tilde d$ are
their superpartners. $D_Z = M_Z^2 \cos\! 2\beta$ where
$\tan\!\beta =\langle H_2^0\rangle / \langle H_{1}^{0} \rangle$
is the ratio of vacuum
expectation values of the two Higgs doublets, $m_{\tilde d_{L,R}}$
are the soft squark masses, $A_d$  are the soft SUSY breaking parameters
describing the strength of the trilinear scalar interactions, and $\mu$
is the supersymmetric Higgs(ino) mass parameter. Once squark
mixing is included, the current eigenstates $\tilde{d}_L, \tilde{d}_R$
become superpositions of the mass eigenstates $\tilde{d}_i$ with
the masses $m_{\tilde d_i}$ and the corresponding  mixing angle
$\theta^d$ defined as
\begin{eqnarray}\label{mixing}
m^2_{\tilde d_{1,2}} = ~~~~~~~~~~~~~~~~~~~~~~~~~~~~~~\nonumber\\
\frac{1}{2}
\left[m^2_{LL} + m^2_{RR}\mp \sqrt{(m^2_{LL}-m^2_{RR})^2 + 4 m^4_{LR}}
\ \right], \\
\label{(mixangle)} \sin\! 2\theta^d = \frac{2 m^2_{(d)LR}}
{m^2_{\tilde{d}_1}-m^2_{\tilde{d}_2}},~~~~~~~~~~~~~
\end{eqnarray}
where $m^2_{LR}, m^2_{LL}, m^2_{RR}$ denote the $(1,2), (1,1), (2,2)$
entries of the mass matrix (\ref{e4}).

The squark-neutrino mechanism is a long-range mechanism involving both heavy squark and light neutrino virtual states as shown in Fig. 1(a).
The bottom part of the diagram corresponds to the ordinary SM charged current interaction.
The $\tilde q_L-\tilde q_R$ - mixing results in lepton number violation with $\Delta L =2$ through the \rp interactions
proportional to the $\lambda^{\prime}\lambda^{\prime}$-coupling. Without the $\tilde q_L-\tilde q_R$ - mixing
the \rp-contribution would be proportional to the $\lambda^{\prime}\lambda^{\prime *}$-coupling, conserving lepton number. Thus, in the latter case
LNV with $\Delta L =2$ is introduced by the Majorana mass of the neutrino with the amplitude of \dbd-decay proportional to
the small neutrino mass and, therefore, is strongly suppressed.

It is straightforward to derive the  effective 4-fermion
$\nu-u-d-e$ vertex induced by the squark
exchange corresponding to the top part of the diagram in Fig. 1(a). The corresponding effective
Lagrangian, after a Fierz rearrangement, takes the form
\begin{eqnarray}\label{L_SUSY}\nn
{\cal L}_{SUSY}^{eff}(x) &=&
\frac{G_F}{8\sqrt{2}} \eta_{(q)LR}^{n1}~ U^{*}_{ni}\times
~~~~~~~~~~~\nonumber\\
&&\left[
\left(\bar \nu_i\ \sigma^{\mu\nu}  (1 + \gamma_5)  e^c\right)
\left(\bar u\   \sigma_{\mu\nu} (1 + \gamma_5)  d\right)
\right. \nonumber\\
&&~+ \left.
2~ \left(\bar \nu_i (1 + \gamma_5) e^c\right)
\left(\bar u (1 + \gamma_5)  d\right)
\right],
\end{eqnarray}
with $\sigma^{\mu\nu} = (i/2)[\gamma^{\mu}, \gamma^{\nu}]$.
Here $U$ is the neutrino mixing matrix
and the SUSY LNV parameter is defined as
\begin{eqnarray}\label{eta}
\eta_{(q)LR}^{nj} &=& \sum_{k} \frac{\lambda'_{j1k}\lambda'_{nk1}}{2
\sqrt{2} G_F}
\sin{2\theta^{d}_{(k)} }\left( \frac{1}{m^2_{\tilde d_1 (k)}} -
\frac{1}{m^2_{\tilde d_2 (k)}}\right).\nonumber\\
\end{eqnarray}
Here we use the notations $d_{(k)} = d, s, b$.
This LNV parameter vanishes in the absence of $\tilde q_L-\tilde q_R$ - mixing when $\theta^{d}=0$, in accordance with the previous comments.

\section{Hadronization prescriptions}

%
To evaluate the contribution of the diagram in Fig. 1(a) to \dbd-decay one first should express the quark fields in terms of
nucleon ones. This procedure, know as hadronization, has so far a rather poor theoretical background. In practice the transformation is carried out
in a phenomenological way, where different possibilities of embedding the quark fields into the interpolating hadronic
fields are considered.

There are the two possibilities \cite{dbd-gluino-neutralino1} of hadronization for
the diagram of Fig. 1(a).

{\bf 1.} The four quark fields can be embedded in the two initial neutrons
and two final protons separately \cite{Moh86,Ver87,dbd-gluino-neutralino}.
This is the conventional two-nucleon (2N) mode of
$0\nu\beta\beta$-decay shown in Fig. 2(b).

{\bf 2.} Another possibility is to incorporate one pair of (anti-)quarks
of the underlying quark-level \rp SUSY transition into one
virtual pion \cite{dbd-gluino-neutralino1,Fedor-Wodecki,ramsey},
while another (anti-)quark pair participates in the charged current SM
transition involving initial neutron and final proton.
In this case, which we call the pion mode of \dbd-decay, the second (the upper one in Fig. 1(c)) nucleon vertex is connected to the leptonic LNV vertex
through the charged pion exchange.
Note that the pion exchange contribution in the charged current SM interaction (the bottom vertex in Fig. 1(c)) is
automatically taken into account by the induced pseudoscalar form factor of the nucleon.

The effective hadronic Lagrangian describing the vertices of the diagrams in Figs. 1(b,c)
can be derived using the method of Ref. \cite{dbd-gluino-neutralino1},
matching the hadronic matrix elements
of the hadronic and the corresponding quark operators. This Lagrangian,
taking into account both the nucleon (p, n) and $\pi$-meson degrees of
freedom in a nucleus, has the following form:
\begin{equation}\label{eff-L}
  {\cal L}_{h} =
{\cal L}_{\beta} + {\cal L}^{PS}_{\rpm} + {\cal L}^{PT}_{\rpm} +{\cal L}^{\pi}_{\rpm} + {\cal L}_s
\\
\end{equation}
with
\begin{eqnarray}\label{CCSM}
{\cal L}_{\beta} &=& \frac{G_F}{\sqrt{2}}~
J^{\mu \dagger}_L ~
\sum_i U_{ei} \bar{e} \gamma_\mu (1 - \gamma_5) \nu_i, \\
\label{PS}
{\cal L}^{PS}_{\rpm}&=& \frac{G_F}{4 \sqrt{2}}
\eta^{11}_{(q)LR} J_{PS}
~\sum_i U^*_{ei} \bar{\nu_i}(1+\gamma_5)e^c, \\
\label{PT}
{\cal L}^{PT}_{\rpm}&=& \frac{G_F}{8\sqrt{2}}
\eta^{11}_{(q)LR} J^{\mu\nu}_{PT}
~\sum_i U^*_{ei} \bar{\nu_i}\sigma_{\mu\nu}(1+\gamma_5)e^c, \\
\label{pi}
\label{pi-nu-e}
{\cal L}^{\pi}_{\rpm}&=& i \frac{G_F}{4 \sqrt{2}}
\eta^{11}_{(q)LR}m^2_\pi h_\pi {\bf \pi^-}
~\sum_i U^*_{ei} \bar{\nu_i}(1+\gamma_5)e^c,\\
\label{pi-NN}
{\cal L}_s &=& g_{\pi NN}~ \bar p\ i \gamma_5 n ~ \pi^+.
\end{eqnarray}
Here, ${\cal L}_{\beta}$ is the charged current SM term, ${\cal L}^{PS}_{\rpm}$, ${\cal L}^{PT}_{\rpm}$ and
${\cal L}^{\pi}_{\rpm}$ are the \rp SUSY induced nucleon-electron-neutrino and pion-electron-neutrino  interactions,
respectively. The coefficient $a_\pi$ in Eq. (\ref{pi-nu-e}) is determined by the matrix element \cite{dbd-gluino-neutralino1}
\begin{eqnarray}
\langle 0|\bar u \gamma_5 d |\pi^-\rangle = i \sqrt{2} f_{\pi}
\frac{m_{\pi}^2}{m_u + m_d} \equiv i m_{\pi}^2 h_{\pi},
\end{eqnarray}
where $f_{\pi} = 0.668~ m_{\pi}$.

The Lagrangian term ${\cal L}_s$  in Eq. (\ref{pi-NN}) stands for the standard pion-nucleon interaction with
the coupling $g_{\pi NN} = 13.4\pm 1$ known from experiment.

The nucleon currents are defined as \cite{Adler}
\begin{eqnarray}\label{currents-1}
J^{\mu \dagger}_L &=& <P(p)|\bar u \gamma^{\mu}(1 - \gamma_5) d |N(p')>
\nonumber\\
&=&  \overline{p}  \left[ g_V(q^2) \gamma^\mu
+ i g_M (q^2) \frac{\sigma^{\mu \nu}}{2 m_p} q_\nu \right.\nonumber\\
&&\left. ~~- g_A(q^2) \gamma^\mu\gamma_5 - g_P(q^2) q^\mu \gamma_5 \right] n,\\
J_{PS} &=& <P(p)| \bar u (1+\gamma_5) d|N(p')> \nonumber\\
&=& \bar p \left[F_S^{(3)}(q^2) + F_P^{(3)}(q^2) \gamma_5\right] n,\\  \nn
J_{PT}^{\mu\nu} &=& <P(p)| \bar u \sigma^{\mu \nu}(1 + \gamma_5) d|N(p')>
\nonumber\\
        &=& \bar p \left(J^{\mu\nu} +
\frac{i}{2} \epsilon^{\mu\nu\rho\sigma} J_{\rho\sigma}\right) n,
\label{tensor2t}
\end{eqnarray}
where $m_p$ is the nucleon mass, $q_\mu = (p-p')_\mu$ is the momentum transferred
to the nucleon vertex with $p$ and $p'$ being the four momenta of neutron and
proton, respectively.
The tensor structure is given by
\ba{JJJ}
J^{\mu\nu} &=& T_1^{(3)}(q^2) \sigma^{\mu\nu} + \frac{i T_2^{(3)}(q^2)}{m_P}
\left(\gamma^{\mu} q^{\nu} - \right. \\
&&\left.  \gamma^{\nu} q^{\mu}\right) +
\frac{T_3^{(3)}(q^2)}{m_P^2}
\left(\sigma^{\mu\rho} q_{\rho} q^{\nu} -  \sigma^{\nu\rho} q_{\rho}
q^{\mu}\right).
\nonumber
\ea
%
The nucleon form factors $g_V(q^2)$, $g_M(q^2)$, $g_A(q^2)$,
$g_P(q^2)$ $F_S^{(3)}$, $F_P^{(3)}$, $T_k^{(3)}$ are real functions
of the squared momentum $q^2$ transferred to a nucleon.
For all the form factors we use the following dipole
parameterizations:
\begin{eqnarray}\label{dip}
\frac{g_{V,A,M}(q^2)}{g_{V,A,M}} =
\frac{F_{S,P}^{(3)}(q^2)}{F_{S,P}^{(3)}(0)} =
\frac{T_i^{(3)}(q^2)}{T_i^{(3)}(0)} =
\left(1-\frac{{q}^{~2}}{\Lambda^2_V}\right)^{-2}
\end{eqnarray}
with $\Lambda^2_V = 0.71 ~(GeV)^2$ and the normalization constants:
$g_V = 1$, $g_A = 1.254$, $g_M = (\mu_p-\mu_n) g_V$, $(\mu_p-\mu_n) = 3.70$.
The induced pseudoscalar coupling is given by the
PCAC relation
\begin{equation}
g_P({q}^{~2}) = {2 m_p g_A({q}^{~2})}/({m^2_\pi - {q}^{~2}}).
\end{equation}
For the normalization constants $F_{S,P}^{(3)}(0)$, $T_i^{(3)}(0)$
we use the results of Ref. \cite{Adler} summarized in Table \ref{table.1},
obtained within the quark bag model and the non-relativistic quark model.

\begin{table}[!t]
\begin{center}
\squeezetable
\caption{Normalizations of nucleon form factors at $q^2 = 0$ calculated in the quark bag model (QBM)
and the non-relativistic quark model (NRQM). The table is taken from Ref. \cite{Adler}.}
\label{table.1}
\renewcommand{\arraystretch}{1.6}
\begin{tabular}{cccccc}\hline\hline
Set & $F_S^{(3)}$ & $F_P^{(3)}$ & $T_1^{(3)}$ & $T_2^{(3)}$ & $T_3^{(3)}$\\
\hline
QBM  & 0.48 & 4.41 & 1.38 & -3.30 & -0.62\\
NRQM   & 0.62 & 4.65 & 1.45 & -1.48 & -0.66\\
\hline\hline
\end{tabular}

\end{center}
\end{table}

\begin{table*}[htb]
\begin{center}
\squeezetable
\caption{Nuclear matrix elements (NMEs) of the
squark-neutrino \rp SUSY
mechanism of $0\nu\beta\beta$-decay. The NMEs of the 2N-mode are calculated
for the two cases of the nucleon form factors: Quark Bag Model (QBM) and
Non-Relativistic Quark Model (NRQM). The quantities $M_{2N}$, $M_{\pi}$ are
the 2N and pion mode nuclear matrix elements averaged over small, medium and large model
spaces (see the text) with their variance $\sigma$ given in parentheses.}
\label{table.2}
\renewcommand{\arraystretch}{1.6}
\begin{tabular}{lccccccccccc}\hline\hline
 & \multicolumn{4}{c}{QBM} & \hspace{0.2cm}
& \multicolumn{4}{c}{NRQM} & \hspace{0.2cm} & \\ \cline{2-5} \cline{7-10}
nucl.  & $M^{\tilde{q}}_{VT}$ & $M^{\tilde{q}}_{MT}$ & $M^{\tilde{q}}_{AP}$ & $M^{\tilde{q}}_{2N}$ & &
         $M^{\tilde{q}}_{VT}$ & $M^{\tilde{q}}_{MT}$ & $M^{\tilde{q}}_{AP}$ & $M^{\tilde{q}}_{2N}$ & & $M^{\tilde{q}}_{\pi}$ \\\hline
$^{76}Ge$  & 185.  & -246. & 29.6 &  -20.0 (33.8) &
           & 102.  & -258. & 31.2 &  -113. (25.7) & & 604. (74) \\
$^{100}Mo$ & 220.  & -244. & 33.0 &   22.9 (1.8)  &
           & 121.  & -256. & 34.8 &  -100. (14.3) & & 594. (80) \\
$^{130}Te$ & 179.    & -206. & 28.4 &   6.3. (20.8) &
           & 99.2 & -217. & 29.8 &  -81.6 (17.2) &  & 517. (32)\\ \hline\hline
\end{tabular}
\end{center}
\end{table*}

\section{\dbd \ Half-Life and Nuclear Matrix Elements}

The matrix element of the SQN-mechanism (SQuark-Neutrino-mechanism)
can be calculated according to the
diagrams of Fig. 1(b,c) with the vertices described by the
effective Lagrangian of Eq. (\ref{eff-L}).
The vertices denoted by black blobs originate in
squark exchange and correspond to the terms (\ref{PS})-(\ref{pi}).
The bottom parts of these
diagrams are the standard model charged current
interactions (\ref{CCSM}).
Applying the standard procedure based on the non-relativistic impulse approximation (for details see
\cite{si99}),
it is straightforward to derive the nuclear matrix elements and the corresponding half-life formula for the \dbd-decay in the
$0^+\rightarrow 0^+$ transition channel with two outgoing electrons
in S-wave states.  It takes the form:
\begin{equation}
\frac{1}{T_{1/2}} = G_{01} |{M}^{\tilde q}_{h}|^2
|\eta^{11}_{(q)LR}|^2~,
\end{equation}
where $G_{01}$ is a precisely calculable phase-space factor. Note that
it is equal to the phase-space factor of the classical
light Majorana neutrino exchange mechanism.
%
The nuclear matrix elements ${M}^{\tilde q}_h$ depend on the hadronization mode. We denote the matrix elements for the 2N and pion modes as
${M}^{\tilde q}_{2N}$ and ${M}^{\tilde q}_{\pi}$, respectively.

We derive these nuclear matrix elements up to the order of $1/m_p$ in the
non-relativistic expansion.
They have the following form.

A) {\it The two nucleon mode} results in the full matrix element
\begin{equation}\label{2N-q}
{M}^{\tilde q}_{2N} = M^{\tilde q}_{AP}+M^{\tilde q}_{MT}+ M^{\tilde
q}_{VT}.
\end{equation}
The partial nuclear matrix elements $M^{\tilde q}_{AP}$,
$M^{\tilde q}_{MT}$ and  $M^{\tilde q}_{VT}$
have their origin
in the interference of the axial-vector and pseudoscalar
currents, weak-magnetic and tensor currents, and vector and tensor
currents, respectively. We note that the $M^{\tilde q}_{VT}$
contribution was neglected in previous studies \cite{Pes}.

Using the method of second quantization in the relative coordinates we obtain
\begin{eqnarray}\label{2N-nme1}
M_{AP}^{\tilde q} = <H_{AP-GT}^{\tilde q}(r_{12}) {\bf
\sigma}_{12} +
H_{AP-T}^{\tilde q}(r_{12}) { S}_{12})>,~~~ \\
\label{2N-nme2}
M_{MT}^{\tilde q} = <H_{MT-GT}^{\tilde q}(r_{12}) {\bf
\sigma}_{12} +
H_{MT-T}^{\tilde q}(r_{12}) { S}_{12})>,~~ \\
\label{2N-nme3}
M_{VT}^{\tilde q} = <H_{VT-F}^{\tilde q}(r_{12}) >.
~~~~~~~~~~~~~~~~~~~~~~~~~~~~~~~~~~
\end{eqnarray}

B) {\it The pion mode} nuclear matrix element obtained within the same framework is
\begin{eqnarray}\label{pi-nme}
M^{\tilde q}_{\pi} = <H_{\pi N-GT}^{\tilde q}(r_{12}) {\bf \sigma}_{12}
+ H_{\pi N-T}^{\tilde q}(r_{12}) { S}_{12})>.~~
\end{eqnarray}

In the above formulas we use the notations:
\begin{eqnarray}
{\bf r}_{12} &= &{\bf r}_1-{\bf r}_2, ~~~ r_{12} = |{\bf r}_{12}|,
~~~
\hat{{\bf r}}_{12} = \frac{{\bf r}_{12}}{r_{12}},~\nonumber \\
\sigma_{12} &=& {\vec{ \sigma}}_1\cdot {\vec{ \sigma}}_2 \nonumber\\
S_{12} &=& 3({\vec{ \sigma}}_1\cdot \hat{{\bf r}}_{12})
       ({\vec{\sigma}}_2 \cdot \hat{{\bf r}}_{12})
      - \sigma_{12}.
\end{eqnarray}
Here, ${\bf r}_1$ and ${\bf r}_2$ are the coordinates of the
nucleons undergoing beta decay.
The form of the matrix element $<{\cal O}(1,2)>$ within
the pn-QRPA will be presented in the next section.

The neutrino-exchange potentials in Eqs. (\ref{2N-nme1})-(\ref{2N-nme3}), (\ref{pi-nme}) are defined as
\begin{eqnarray}\label{nu-pot}
H^{\tilde q}_{type-K} (r_{12}) =
\frac{2}{\pi g_A^2} {R}\times~~~~~~~~~~~~~~~
\nonumber\\
\int_0^{\infty}~ f_K(qr_{12})~ \frac{ h^{\tilde q}_{type-K} (q^2) q
dq } {q + E^{m}_{J} - (E^i_{g.s.} + E^f_{g.s.})/2} ~,
\end{eqnarray}
where $type-K$ stands for $VT-F, MT-GT, MT-T, AP-GT, AP-T, \pi N -GT, \pi N -T$. We also define
$f_{F,GT}(qr_{12}) = j_0(q r_{12})$ and $f_{T}(qr_{12}) = -
j_2(qr_{12})$, where $j_{0,2}$  are the spherical Bessel functions.
In the denominator of Eq. (\ref{nu-pot}) $E^i_{g.s.}$ and $E^f_{g.s.}$ are the
ground state energies of the initial and final nuclei, respectively,
while $E^{m}_{J}$ is the energy of the intermediate nuclear states. $R =
r_0 A^{1/3}$ is the mean nuclear radius with $r_0 = 1.1~ fm$.
The functions $h^{\tilde q}_{type-K}$ in Eq. (\ref{nu-pot}) can be expressed in the following form:
\begin{eqnarray}\label{VT-F}
h^{\tilde q}_{VT-F}({\vec q}^{~2} ) &=&
- g_{V}({\vec q}^{~2})
 \left( T^{(3)}_1({\vec q}^{~2})- 2 T^{(3)}_2({\vec q}^{~2})\right)
\nonumber\\
&&\times \frac{{\vec q}^{~2}}{2 m_p m_e},\\
\label{MT-GT}
h^{\tilde q}_{MT-GT}({\vec q}^{~2} ) &=& -\frac{2}{3} \left(g_{V} ({\vec q}^{~2}) + g_{M} ({\vec q}^{~2})\right)
 T^{(3)}_1({\vec q}^{~2}) \nonumber\\
&&\times \frac{{\vec q}^{~2}}{2 m_p m_e},\\
\label{AP-GT}
h^{\tilde q}_{AP-GT}({\vec q}^{~2} ) &=& \frac{1}{6}  g_{A} ({\vec q}^{~2})
 F^{(3)}_P({\vec q}^{~2})\nonumber\\
&&\times \frac{m^2_\pi}{2 m_p m_e} \frac{{\vec q}^{~2}}{{\vec
q}^{~2}+m^2_\pi},
\\
\label{MT-T}
h^{\tilde q}_{MT-T}({\vec q}^{~2} ) &=& -\frac{1}{2} h^{\tilde
q}_{MT-GT}({\vec q}^{~2} ),\nonumber\\
h^{\tilde q}_{AP-T}({\vec q}^{~2}) &=& h^{\tilde q}_{AP-GT}({\vec q}^{~2} ),\\
\label{pi N-GT}
h^{\tilde q}_{\pi N-GT}({\vec q}^{~2} ) &=& - \frac{1}{6}~ g^2_{A} ({\vec q}^{~2}) \frac{m^4_\pi}{m_e (m_u + m_d)}\nonumber\\
&&\times \frac{{\vec q}^{~2}}{({\vec q}^{~2}+m^2_\pi)^2},\\
\label{pi N-T}
h^{\tilde q}_{\pi N-T}({\vec q}^{~2} ) &=& h^{\tilde q}_{\pi N-GT}({\vec q}^{~2} ),
\end{eqnarray}
where the nucleon form factors $g_K$ ($K=V,~A$ and $M$),  $F^{(3)}$ and $T^{(3)}$ are defined in sect. III.
In the case of the pion-exchange mechanism, Eq. (\ref{pi N-GT}), we used
the Goldberger-Treimann relation \cite{goldtr}.

The following comment is in order. Within the considered non-relativistic approximation the NMEs of the 2N-mode, Eqs (\ref{VT-F})-(\ref{MT-T}),
depend nearly on all the nucleon form factors except for $F^{(3)}_S$ and $T_3^{(3)}$.
As seen from Table \ref{table.1} these form factors
are rather model dependent quantities.
On the other hand the pion-mode NMEs, Eqs. (\ref{pi N-GT})-(\ref{pi N-T}),
involve only the axial-vector form factor $g_A$, which is
well-known from experimental measurements.
Thus,  the pion-mode NMEs are expected to be significantly less dependent on
nucleon structure models than the NMEs  of the  2N-mode.

\section{Calculation of Nuclear Matrix Elements}

We calculate the NMEs introduced
in the previous section. The partial matrix
elements in Eqs. (\ref{2N-nme1})-(\ref{pi-nme})
are given by the expression \cite{si99}:
\begin{eqnarray}\label{eq:long}
M_K  =  \sum_{J^{\pi},k_i,k_f,\mathcal{J}} \sum_{pnp'n'}
(-1)^{j_n + j_{p'} + J + {\mathcal J}} \times~~~~~~~~~~
\\
\sqrt{ 2 {\mathcal J} + 1}
\left\{
\begin{array}{c c c}
j_p & j_n & J  \\
 j_{n'} & j_{p'} & {\mathcal J}
\end{array}
\right\}  \times~~~~~~~~~~~~~~~~~~~
\nonumber \\
\langle p(1), p'(2); {\mathcal J} \parallel f(r_{12})
O_K f(r_{12}) \parallel n(1), n'(2); {\mathcal J} \rangle \times~~
\nonumber \\
\langle 0_f^+ ||
[ \widetilde{c_{p'}^+ \tilde{c}_{n'}}]_J || J^{\pi} k_f \rangle
\langle  J^{\pi} k_f |  J^{\pi} k_i \rangle
 \langle  J^{\pi} k_fi|| [c_p^+ \tilde{c}_n]_J || 0_i^+ \rangle ~.
\nonumber
\end{eqnarray}
The sum over $J^{\pi}$ represents the summation over the
states of the  intermediate odd-odd nucleus with angular momentum
$J$, parity $\pi$ and energies $E^{k_i,k_f}_{J^\pi}$.
The indices $k_i$ and $k_f$ represent states which are calculated
from the ground state of the initial $(i)$ and final $(f)$ nuclei,
respectively. The
overlap factor $\langle  J^{\pi} k_f |  J^{\pi} k_i \rangle$
accounts for the difference between them.
The reduced matrix elements of the one-body operators
$c_p^+ \tilde{c}_n$ ($\tilde{c}_n$ denotes the time-reversed state)
in (\ref{eq:long}) depend on the BCS occupation
coefficients $u_i,v_j$ and on the QRPA vectors $X,Y$ \cite{si99}.

The operators $O_K$,with $K = F,GT,T$, in (\ref{eq:long}) depend on
the distance $r_{12}$ between two inicial neutrons that are transferred
into two protons, the relevant spin and isospin operators as well
as on the energies of the excited states $E^{k_i,k_f}_{J^\pi}$,
albeit in practice rather weakly. Short range
correlations of the two initial neutrons and the two final protons are
described by the Jastrow function $f(r_{12})$
according to Ref. \cite{MS1976}.

We apply the pn-QRPA to calculate the $0\nu\beta\beta$-decay
NMEs for $^{76}Ge$, $^{100}Mo$ and $^{130}Te$. For these nuclei
strong experimental limits on the $0\nu\beta\beta$-decay half-life
were found [see Eqs. (\ref{H-M}) and (\ref{nemo-cuore})]
and these nuclei are also considered as candidate
sources for the next generation of the  $0\nu\beta\beta$-decay
experiments. For each of them three choices of
single particle basis (minimal, intermediate and large
s.p. model spaces) are considered according to Ref.
\cite{FedorVogel}. The single particle energies are
obtained from a spherical, Coulomb corrected Woods-Saxon
potential. The Brueckner G-matrices of the Bonn potential
are used as a two-body interaction. The pairing interactions
of the nuclear Hamiltonian $H$ are adjusted to fit the
empirical pairing gaps. In addition,  we renormalize the
particle-particle and particle-hole channels of $H$
by introducing the parameters $g_{pp}$ and $g_{ph}$, respectively.
In the calculation we use $g_{ph}=1.0$ and  $g_{pp}$ is fixed
so that the known half-life of the $2\nu\beta\beta$-decay
is correctly reproduced for each model space \cite{rodin}.
Recently, it was found that such a procedure makes the $0\nu\beta\beta$-decay
NMEs essentially independent of the size of the single particle
basis and the nuclear structure input \cite{FedorVogel,rodin}.
This procedure allowed us also to evaluate the uncertainties in the
calculated NMEs.

Numerical results of the NME calculation are summarized in Table \ref{table.2}.
The presented values of the partial nuclear matrix elements
$M^{\tilde q}_{VT}$, $M^{\tilde q}_{MT}$
and $M^{\tilde q}_{AP}$ have been obtained for  the intermediate s.p.
 model  space: 12 levels for A=76, 16 levels for A=100, and 18
levels for A=130.
The values of  NMEs for the three studied isotopes
are comparable in magnitude. We have found  that the considered two-nucleon
short range correlations suppress the  pion-mode NME by about $20\%$.
We recall that the similar reduction occurs also for the NMEs of the ``classic"
light Majorana neutrino mass mechanism of the $0\nu\beta\beta$-decay.

As seen from Table \ref{table.2}, the pion-mode NMEs
are from 6 to 30 times larger, depending on the nucleon structure model, than the corresponding
2N-mode NMEs for all the studied nuclei.
This allows us to conclude that
in the squark-neutrino \rp SUSY mechanism of \dbd-decay
the pion-mode of hadronization clearly dominates over the 2N-mode.

Let us point out that in the $2N$-mode NMEs there is
a strong cancelation between the $M^{\tilde q}_{VT}$ and
$M^{\tilde q}_{MT}$ partial matrix elements (see Table \ref{table.2})
originating from the tensor nucleon current (\ref{tensor2t}).
As a result despite the individual values of both matrix elements
$M^{\tilde q}_{VT}$ and $M^{\tilde q}_{MT}$
are significantly larger than the pseudoscalar contribution $M^{\tilde q}_{AP}$ it turns
out that their cumulative effect, $M^{\tilde q}_{VT} + M^{\tilde q}_{MT}$, is suppressed and can be either comparable in magnitude
with $M^{\tilde q}_{AP}$ or larger than it, depending on the model of nucleon structure (QBM or NRQM).
At this point we disagree with Ref. \cite{Pes} claiming that the tensor contribution is always significantly larger than the pseudoscalar one.
This disagreement can be accounted for the fact that the contribution $M^{\tilde q}_{VT}$ was missed in Ref. \cite{Pes} leaving the large term,
analogous to our $M^{\tilde q}_{MT}$, uncompensated. As a result the tensor contribution, represented by only this term, becomes
significantly larger than the pseudoscalar contribution for both QBM and NRQM models of nucleon form factors. As we have shown this is not the case when
all the contributions of the tensor nucleon current (\ref{tensor2t}) are properly taken into account.

In Table \ref{table.2} the NMEs of the pion and $2N$ modes
are given with their variances $\sigma$ (evaluated according
to Ref. \cite{FedorVogel}) due to uncertainties of nuclear structure input.
In both cases these variances are within $10\%$.
This comparatively low uncertainty was obtained mainly due to the procedure
we chose for fixing the
residual interaction of the nuclear Hamiltonian which includes the calculation
with three model spaces \cite{FedorVogel,rodin}.
As to the uncertainties of the nucleon form factors,
from Table \ref{table.2} it is evident that they are very significant for the 2N-mode NMEs.
For the two models of nucleon form factors considered here, the Quark Bag Model and the
Non-Relativistic Quark Model, the values of the partial NMEs of the 2N-mode change up to a factor 2 while the total 2N-mode NMEs up to an order of magnitude.
Fortunately, as we commented at the end of Sect. IV the dominant pion-mode contribution is free of the nucleon structure uncertainties
making their impact on the resulting NMEs insignificant.

\begin{table}[t]
\begin{center}
\squeezetable
\caption{Upper bounds on the \rp SUSY parameter $\eta^{11}_{(q)LR}$
as well as
on the related products of the trilinear \rp-couplings
$\lambda^{\prime}_{11k}\lambda^{\prime}_{1k1}$ (k=1,2,3) for $\Lambda_{SUSY}=100$ GeV (see scaling law in Eq. (\ref{lim})) deduced from the
current lower bounds on the half-life of $0\nu\beta\beta$-decay for $^{76}Ge$,
$^{100}Mo$ and $^{130}Te$.
}
\label{table.3}
\renewcommand{\arraystretch}{1.6}
\begin{tabular}{lcccccc}\hline\hline
nucl. & $T^{0\nu-exp}_{1/2}$ [Ref.] & $\eta^{11}_{(q)LR}$ & & $\lambda^{'}_{111}\lambda^{'}_{111}$ & $\lambda^{'}_{112}\lambda^{'}_{121}$ &$\lambda^{'}_{113}\lambda^{'}_{131}$\\
 & (years)\hspace{0.9cm} & & & & &   \\ \hline
$^{76}Ge$  & $\geq 1.9~10^{25}$ \cite{ge76} & $4.3~ 10^{-9}$ &
           & $7.7~10^{-6}$ & $4.0~10^{-7}$ & $1.7~10^{-8}$ \\
$^{100}Mo$ & $\geq 5.8~10^{23}$ \cite{mo100} & $9.2~ 10^{-9}$ &
           & $1.7~10^{-5}$ & $8.7~10^{-7}$ & $3.6~10^{-8}$ \\
$^{130}Te$ &$\geq 3.0~10^{24}$  \cite{te130} & $4.7~ 10^{-9}$ &
           & $8.5~10^{-6}$ & $4.5~10^{-7}$ & $1.9~10^{-8}$ \\
\hline\hline
\end{tabular}
\end{center}
\end{table}


\section{Discussion and Conclusions}

With the values of the nuclear matrix elements of the dominant pion-mode contribution given in Table \ref{table.2} we derived
the experimental upper limits on the LNV parameter $\eta^{11}_{(q)LR}$ introduced in Eq. (\ref{L_SUSY}).
Then, using Eq. (\ref{eta}), we extracted upper limits on the products of the trilinear \rp-couplings $\lambda^{'}_{11k}\lambda^{'}_{1k1}$ (k=1,2,3). These limits,
presented in Table \ref{table.3},
have been derived under the conventional simplifying assumptions. We assumed all the squark masses and the trilinear soft SUSY breaking parameters $A_d$
to be approximately equal to a common SUSY breaking scale $\Lambda_{SUSY}$. Thus we approximately write
\begin{eqnarray}\label{lim}
\lambda^{\prime}_{11k} \lambda^{\prime}_{1k1} \leq \epsilon_k
\frac{1}{\sqrt{T^{0\nu - exp}_{1/2} G_{01}}}
\frac{1}{M^{\tilde{q}}_\pi} \left(\frac{\Lambda_{SUSY}}{100\mbox{GeV}}\right)^3
\end{eqnarray}
with $\epsilon_k=(1.8\times 10^3; 94.2; 3.9)$ calculated for the current quark masses $m_d = 9$ MeV, $m_s=175$ MeV and $m_b=4.2$~GeV respectively.
In the above formula $T^{0\nu-exp}_{1/2}$ is the experimental lower bound on the half-life of  a certain nucleus.
In Table \ref{table.3} we show the upper limits on the \rp SUSY parameters,
extracted from the existing experimental lower bounds on the half-lives of the three nuclei ${}^{76}Ge$, ${}^{100}Mo$ and ${}^{130}Te$.
Our limit for ${}^{76}Ge$ is about an order of magnitude better than the limit previously obtained in Ref. \cite{HKK:96, Pes}  on the basis of the
2N-mode of hadronization taking into account only the pseudoscalar current contribution, $M^{\tilde q}_{AP}$.
%
%

In conclusion, we analyzed \dbd-decay of several nuclei induced by the LNV effective operators originating from the \rp SUSY trilinear interactions
involving squark and neutrino exchange.  We focussed on the hadronization prescription of the quark-level operators and analyzed both the conventional 2N-mode and
the pion-mode of hadronization. We have shown that the pion-mode absolutely dominates over the 2N-mode. Previously, in Refs. \cite{dbd-gluino-neutralino1} it was
demonstrated that the pion-mode dominates over the 2N-mode in the case of the short-range \rp SUSY mechanism. Thus, with the result of the present paper
we conclude that all the mechanisms based on the trilinear \rp SUSY interactions dominantly contribute to \dbd-decay via the pion-mode of hadronization.

\section{Acknowledgments}

We are thankful to M. Hirsch for valuable discussions.
This work was supported in part by CONICYT (Chile) under grant PBCT/No.285/2006,
the EU Integrated Infrastructure Initiative Hadronphysics project
under the contract  RII3-CT-2004-506078, the EU ILIAS project under the contract
RII3-CT-2004-506222, the DFG project 436 SLK 17/298, the Transregio
Project TR27 "Neutrinos and Beyond" and by the VEGA Grant agency
under the contract No.~1/0249/03.

\end{document}